\documentclass[epsfig,12pt]{article}
\usepackage{makeidx}
\usepackage{amsmath}
\usepackage{amsfonts}
\usepackage{amssymb}
\usepackage{graphicx}

\input epsf.sty
\makeatletter \@addtoreset{equation}{section}
\renewcommand{\theequation}{\thesection.\arabic{equation}}
\input{tcilatex}
\begin{document}

\title{\vspace{-2cm}%
\rightline{\mbox{\small
{LPHE Preprint 2012}} \vspace {1cm}} \textbf{Weak Coupling Chambers in} $%
\mathcal{N}=2$ \textbf{BPS Quiver Theory}}
\author{El Hassan Saidi\thanks{%
E-mail: h-saidi@fsr.ac.ma}\  \and \  \\
{\small Lab Of High Energy Physics, Modeling and Simulations, Faculty of
Science, }\\
{\small University Mohammed V-Agdal, 4 Avenue Ibn Battota, }\\
{\small Rabat, Morocco.}\\
and\\
{\small Centre Of Physics and Mathematics, CPM-CNESTEN, Rabat, Morocco}}
\maketitle

\begin{abstract}
Using recent results on BPS quiver theory, we develop a group theoretical
method to describe the quiver mutations encoding the quantum mechanical
duality relating the spectra of distinct quivers. We illustrate the method
by computing the BPS spectrum of the infinite weak chamber of some examples
of $\mathcal{N}=2$ supersymmetric gauge models without and with quark
hypermultiplets. \newline
\end{abstract}

\tableofcontents

\section{Introduction}

Recently a BPS quiver theory has been proposed in \textrm{\cite{1A,1B}} to
build the full set of BPS spectra in 4D $\mathcal{N}=2$ supersymmetric
quantum field theory (QFT$_{4}$). This quiver model is based on quantum
mechanical dualities encoded by \emph{quiver mutations} and relating
distinct quivers of the theory. \newline
In this paper, we develop the group structure property\textrm{\footnote{%
the set quiver mutations for higher dimensional gauge symmetries has a
groupoid structure \cite{3B}.}} of the set quiver mutations in $\mathcal{N}=2
$ supersymmetric quantum theory with SU$\left( 2\right) $ gauge symmetry;
and use this remarkable property to approach the complexity of the infinite
BPS weak coupling chambers of these supersymmetric theories. To illustrate
the key idea of our method, we consider below the weak coupling chambers of
3 particular supersymmetric QFTs: $\left( i\right) $ the weak coupling
chamber of the $\mathcal{N}=2$ supersymmetric pure SU$\left( 2\right) $
gauge theory spontaneously broken down to $U\left( 1\right) $. The BPS
spectrum of this model has been studied in \textrm{\cite{1A,1B}; }see also%
\textrm{\  \cite{1C} }and refs therein\textrm{; }but here we will re-derive
this infinite spectrum from a group theory approach\textrm{. }$\left(
ii\right) $ the weak coupling chambers of the $\mathcal{N}=2$ supersymmetric
SU$\left( 2\right) $ gauge theory with one quark hypermultiplet; and $\left(
iii\right) $ with two quark hypermultiplets. We end this paper by a
conclusion and a comment on the method including the link between the quiver
mutation symmetries and the strong/weak coupling chambers of $\mathcal{N}=2$
QFT$_{4}$s.

\section{Weak coupling chamber of $SU\left( 2\right) $ model}

We begin by recalling that the low energy $\mathcal{N}=2$ supersymmetric $%
SU\left( 2\right) $ QFT has a monopole $\mathfrak{M}$ and a dyon $\mathfrak{D%
}$ believed to be two elementary BPS states of the $\mathcal{N}=2$
superalgebra with central charges \textrm{\cite{2A,2B,2C,2D}; see also \cite%
{3A,A3,3} for a review}. Let denote by $X$ and $Y$ the complex central
charges of these two BPS states; and by $\gamma _{1}$ and $\gamma _{2}$
their electric magnetic (EM) vectors. These central charges and EM charge
vectors are related; and play an important role in the study of the BPS
spectra of this supersymmetric gauge theory; the masses of $\mathfrak{M}$
and $\mathfrak{D}$ are proportional the absolute values $\left \vert
X\right
\vert $ and $\left \vert Y\right \vert $; and the arguments $\arg X$%
, $\arg Y $ for distinguishing the two possible BPS chambers of the $%
SU\left( 2\right) $ gauge theory namely: $\left( a\right) $ $\arg X<\arg Y$
describing the strong coupling chamber of the supersymmetric $SU\left(
2\right) $ QFT; and $\left( b\right) $ $\arg X>\arg Y$ for the weak coupling
one. The first one is finite as it contains the two elementary BPS states
and their CPT conjugates. The second BPS chamber is very rich; it is
infinite and remarkably built in the framework of BPS quiver theory that we
consider here below.

\subsection{BPS spectrum}

This spectrum has been explicitly built in \textrm{\cite{1A,1B} }by using
the BPS quiver theory. There, the BPS states of the weak coupling limit of
the\textrm{\ }$\mathcal{N}=2$ supersymmetric $SU\left( 2\right) $ QFT is
obtained from the primitive quiver $\mathfrak{Q}_{0}$ by performing two
kinds of quiver mutations: left mutations and right ones. Under left
mutations of $\mathfrak{Q}_{0}$, made of the monopole $\mathfrak{M}$ and the
dyon $\mathfrak{D}$, we get an infinite set of BPS and anti-BPS states with
respective electric-magnetic (EM) charges $\gamma _{left}^{\left( n\right) }$%
, $\gamma _{left}^{\prime \left( n\right) }$ as follows%
\begin{eqnarray}
&&%
\begin{tabular}{ll}
$\gamma _{left}^{\left( n\right) }=$ & $\left( n+1\right) \gamma
_{1}+n\gamma _{2}$ \\ 
$\gamma _{left}^{\prime \left( n\right) }=$ & $-n\gamma _{1}-\left(
n+1\right) \gamma _{2}$%
\end{tabular}
\label{l} \\
&&  \notag
\end{eqnarray}%
with $n$ an arbitrary positive integer; $n\geq 0$. Under the right mutation,
we also obtain an infinite BPS and anti-BPS states with EM charges $\gamma
_{right}^{\left( n\right) }$, $\gamma _{right}^{\prime \left( n\right) }$
given by,%
\begin{eqnarray}
&&%
\begin{tabular}{ll}
$\gamma _{right}^{\left( n\right) }=$ & $n\gamma _{1}+\left( n+1\right)
\gamma _{2}$ \\ 
$\gamma _{right}^{\left( n\right) }=$ & $-\left( n+1\right) \gamma
_{1}-n\gamma _{2}$%
\end{tabular}
\label{m} \\
&&  \notag
\end{eqnarray}%
The left and right spectra are related by interchanging the role of the $%
\gamma _{1}$ and $\gamma _{2}$; or by performing $\gamma _{i}\rightarrow
-\gamma _{i}$; the second link may be also interpreted in terms of CPT
invariance of the BPS chamber. \newline
Below we show that this spectrum can be also derived by using a group theory
method; this approach has the property of giving a group theoretical
interpretation of quiver mutations; in particular the right mutations are
precisely the inverse of the left ones. It also has the power to deal with
the complexity of weak chamber of supersymmetric gauge theories without and
with matter including higher dimensional gauge symmetries.

\subsection{Group theory approach}

The key idea of our group theoretical approach that we develop below relies
on the two following basic things: (\textbf{1}) think about the primitive
quiver $\mathfrak{Q}_{0}$, of the BPS quiver theory of $\mathcal{N}=2$
supersymmetric SU$\left( 2\right) $ gauge model, as given by the pair $%
\left( \Upsilon ^{0},\mathcal{A}^{0}\right) $ consisting of a vector $%
\Upsilon ^{0}$ and an intersection matrix $\mathcal{J}^{0}$ as follows 
\begin{equation}
\begin{tabular}{llll}
$\Upsilon ^{0}=\left( 
\begin{array}{c}
\gamma _{1} \\ 
\gamma _{2}%
\end{array}%
\right) $ & , & $\mathcal{J}^{0}=\Upsilon ^{0}\circ \Upsilon ^{0}$ & 
\end{tabular}
\label{Q0}
\end{equation}%
with $\Upsilon \circ \Upsilon $ standing for the electric-magnetic product. (%
\textbf{2}) realize the quiver mutations of the primitive $\mathfrak{Q}_{0}$
in terms of a set $\left \{ \mathcal{M}_{m}\right \} $ of discrete rotations
in the space of the EM charges as follows%
\begin{equation}
\begin{tabular}{llll}
$\Upsilon ^{\left( m\right) }=\mathcal{M}_{m}\Upsilon ^{0}\ $ & , & $%
\mathcal{J}^{\left( m\right) }=\mathcal{M}_{m}\mathcal{J}^{0}\mathcal{M}%
_{m}^{T}$ & 
\end{tabular}
\label{QM}
\end{equation}%
with $m$ an arbitrary positive integer. In this view, a generic mutation $%
\mathcal{M}_{m}$ maps the primitive $\mathfrak{Q}_{0}$ into a mutated one $%
\mathfrak{Q}_{m}$ given by a pair $\left( \Upsilon ^{\left( m\right) },%
\mathcal{J}^{\left( m\right) }\right) $. In the above relations, the
mutation transformations are given by invertible 2$\times $2 matrices with
integer entries; and to compare with the result of \textrm{\cite{1A,1B}},
these $\mathcal{M}_{m}$'s have to encode both the left $\mathcal{M}%
_{m}^{left}$ and right $\mathcal{M}_{m}^{right}$ mutations of (\ref{l}-\ref%
{m}). To that purpose, let us refer to the matrices $\mathcal{M}_{m}^{left}$
and $\mathcal{M}_{m}^{right}$ respectively as $\mathcal{L}_{m}$ and $%
\mathcal{R}_{m}$; and the set of mutations of the weak chamber of this BPS
quiver theory as%
\begin{equation}
\mathcal{G}_{weak}^{su_{2}}=\left \{ \mathcal{L}_{m},\text{\ }\mathcal{R}%
_{m};\text{ }m\in \mathbb{N}\right \} \subseteq GL\left( 2,\mathbb{Z}\right)
\label{w}
\end{equation}%
From the explicit computation of the $\mathcal{L}_{m}$\ and\ right $\mathcal{%
R}_{m}$ mutations, we learn that their expressions depend on the parity of
the positive integer $m$. This is why it is helpful to split the $\mathcal{L}%
_{m}$'s and the $\mathcal{R}_{m}$'s using even and odd integers like%
\begin{equation}
\begin{tabular}{lll}
$\mathcal{L}_{m}=\left( \mathcal{L}_{2k},\mathcal{L}_{2k+1}\right) $ & $,%
\text{\  \ }$ & $\mathcal{R}_{m}=\left( \mathcal{R}_{2k},\mathcal{R}%
_{2k+1}\right) $%
\end{tabular}
\label{LR}
\end{equation}%
with $k$ an arbitrary positive integer. Straightforward calculations using
quiver mutation rules of the quiver BPS theory show that the explicit
expressions of the mutations matrices $\mathcal{L}_{2k}$, $\mathcal{L}%
_{2k+1} $ as well as $\mathcal{R}_{2k}$, $\mathcal{R}_{2k+1}$ are as follows:%
\begin{eqnarray}
&&%
\begin{tabular}{l|l}
\hline
\  \  \  \  \  \  \  \  \  \  \  \  \  \  \  \  \ left sector & \  \  \  \  \  \  \  \  \  \  \  \  \  \
\  \  \ right sector \\ 
$%
\begin{tabular}{lll}
$\mathcal{L}_{2k}$ & $=$ & $\left( 
\begin{array}{cc}
1+2k & 2k \\ 
-2k & 1-2k%
\end{array}%
\right) $ \\ 
&  &  \\ 
$\mathcal{L}_{2k+1}$ & $=$ & $\left( 
\begin{array}{cc}
-1-2k & -2k \\ 
2k+2 & 1+2k%
\end{array}%
\right) $ \\ 
&  & 
\end{tabular}%
$ & $%
\begin{tabular}{lll}
$\mathcal{R}_{2k}$ & $=$ & $\left( 
\begin{array}{cc}
1-2k & -2k \\ 
2k & 1+2k%
\end{array}%
\right) $ \\ 
&  &  \\ 
$\mathcal{R}_{2k+1}$ & $=$ & $\left( 
\begin{array}{cc}
2k+1 & 2k+2 \\ 
-2k & -2k-1%
\end{array}%
\right) $ \\ 
&  & 
\end{tabular}%
$ \\ \hline
\end{tabular}
\label{REP} \\
&&  \notag
\end{eqnarray}%
These mutation matrices obey a set of remarkable properties such as the
involutions $\left( \mathcal{L}_{2k+1}\right) ^{2}=$ $I_{id}$ and $\  \left( 
\mathcal{R}_{2k+1}\right) ^{2}=$ $I_{id},$\ for any integer $k$; and
moreover 
\begin{equation}
\begin{tabular}{ll}
$\det \mathcal{L}_{2k}$ & $=\det \mathcal{R}_{2k}=+1,$ $\  \  \forall k$ \\ 
$\det \mathcal{L}_{2k+1}$ & $=\det \mathcal{R}_{2k+1}=-1,$ $\  \  \forall k$%
\end{tabular}%
\end{equation}%
showing that $\mathcal{L}_{2k}$, $\mathcal{L}_{2k+1}$,$\mathcal{R}_{2k}$, $%
\mathcal{R}_{2k+1}$ are invertible matrices; and therefore the set $\mathcal{%
G}_{weak}^{su_{2}}$ form a subgroup of $GL\left( 2,\mathbb{Z}\right) $. To
establish the relations (\ref{REP}), one uses the representation%
\begin{equation}
\begin{tabular}{lll}
$\mathcal{L}_{2k}=\left( BA\right) ^{k}$ & , & $\mathcal{R}_{2k}=\left(
AB\right) ^{k}$ \\ 
&  &  \\ 
$\mathcal{L}_{2k+1}=A\mathcal{L}_{2k}$ & , & $\mathcal{R}_{2k+1}=\mathcal{R}%
_{2k}A$%
\end{tabular}
\label{ba}
\end{equation}%
showing that the infinite set $\mathcal{G}_{weak}^{su_{2}}$ of quiver
mutations is indeed a subgroup of $GL\left( 2,\mathbb{Z}\right) $. This set
is generated by two reflections $A$ and $B$ given by the triangular matrices 
\begin{equation}
\begin{tabular}{lll}
$A=\left( 
\begin{array}{cc}
-1 & 0 \\ 
2 & 1%
\end{array}%
\right) $ & , & $B=\left( 
\begin{array}{cc}
1 & 2 \\ 
0 & -1%
\end{array}%
\right) $%
\end{tabular}
\label{AB}
\end{equation}%
satisfying the usual reflection property $A^{2}=B^{2}=I_{id}$ and allow to
build all possible elements of $\mathcal{G}_{weak}^{su_{2}}$ as in (\ref{ba}%
). Our way of doing gives then a group theoretical realization of the
quantum mechanical duality of the BPS quiver theory. The $\mathcal{G}%
_{weak}^{su_{2}}$ group permits to map the primitive quiver $\mathfrak{Q}%
_{0} $, thought of as in (\ref{Q0}), to any dual quiver $\mathfrak{Q}_{m}$
given by (\ref{QM}). It allows to get directly the BPS spectrum of the weak
chamber since the EM charge vectors of the BPS states are precisely given by
the row of the matrices (\ref{REP}). We have:%
\begin{eqnarray}
&&%
\begin{tabular}{l|l|ll}
& \  \  \  \  \  \  \ left sector & \multicolumn{2}{|l}{\  \  \  \  \  \  \ right sector}
\\ \hline
$\gamma _{1}^{\left( 2k\right) }$ & $=\gamma _{1}+2kw$ & $\gamma
_{1}^{\left( 2k\right) }$ & $=\gamma _{1}-2kw$ \\ 
&  &  &  \\ 
$\gamma _{2}^{\left( 2k\right) }$ & $=\gamma _{2}-2kw$ & $\gamma
_{2}^{\left( 2k\right) }$ & $=\gamma _{2}+2kw$ \\ 
&  &  &  \\ 
$\gamma _{1}^{\left( 2k+1\right) }$ & $=\gamma _{2}-\left( 2k+1\right) w$ & $%
\gamma _{1}^{\left( 2k+1\right) }$ & $=\gamma _{2}+\left( 2k+1\right) w$ \\ 
&  &  &  \\ 
$\gamma _{2}^{\left( 2k+1\right) }$ & $=\gamma _{1}+\left( 2k+1\right) w$ & $%
\gamma _{2}^{\left( 2k+1\right) }$ & ${\normalsize =\gamma _{1}-}\left(
2k+1\right) {\normalsize w}$%
\end{tabular}
\label{2K} \\
&&  \notag
\end{eqnarray}%
with $w=\gamma _{1}+\gamma _{2}$ giving the EM charge of the W-boson vector
particle. From the above results, we learn that the BPS spectrum of the weak
coupling chamber of the supersymmetric SU$\left( 2\right) $ gauge theory is
infinite; and moreover the number of BPS states grows linearly with the
positive integer $k$.\ If defining the asymptotic limit of the mutation
matrices $\mathcal{M}_{n}$ and by the regularized relation $\mathcal{M}%
_{\infty }=\lim_{n\rightarrow \infty }\left( \frac{1}{n}\mathcal{M}%
_{n}\right) $ we find%
\begin{equation}
\begin{tabular}{lll}
$\mathcal{M}_{\infty }=\left( 
\begin{array}{cc}
1 & 1 \\ 
-1 & -1%
\end{array}%
\right) $ & , & $A\mathcal{M}_{\infty }=\left( 
\begin{array}{cc}
-1 & -1 \\ 
1 & 1%
\end{array}%
\right) $%
\end{tabular}%
\end{equation}%
with $\det \mathcal{M}_{\infty }=0$ and similarly $\det A\mathcal{M}_{\infty
}=0$. These singular asymptotic limits describe the BPS particles with
charges $\pm w$. These are vector particles having only electric charges;
but no magnetic ones; they are associated with $\mathcal{N}=1$ massive $%
W^{\pm }$ vector multiplets.

\section{Adding a hypermultiplet}

In this section, we extend the above $\mathcal{N}=2$ supersymmetric $%
SU\left( 2\right) $ gauge model to include a quark hypermultiplet with a
unit flavor charge under the $U_{f}\left( 1\right) $ symmetry; and use the
mutation group method we have developed above to compute the BPS spectrum of
the weak coupling chamber of this theory. The elementary BPS particles are
then given by the monopole $\mathfrak{M}$, the dyon $\mathfrak{D}$ and the
quark hypermultiplet $\mathfrak{H}$ with respective electric-magnetic
charges $\gamma _{1},$ $\gamma _{2},$ $\gamma _{3}$ as follows%
\begin{equation}
\begin{tabular}{lll}
$\gamma _{1}=\left( 
\begin{array}{c}
0 \\ 
1 \\ 
\frac{1}{2}%
\end{array}%
\right) ,$ & $\gamma _{2}=\left( 
\begin{array}{c}
1 \\ 
-1 \\ 
\frac{1}{2}%
\end{array}%
\right) ,$ & $\gamma _{3}=\left( 
\begin{array}{c}
1 \\ 
0 \\ 
-1%
\end{array}%
\right) $%
\end{tabular}
\label{EM}
\end{equation}%
The third entry of these vectors is the charge under the $U_{f}\left(
1\right) \simeq SO_{f}\left( 2\right) $ flavor symmetry; see also \textrm{%
\cite{1B} and \cite{1C}} for details regarding the obtention of the
primitive quiver $\mathfrak{Q}_{0}\left( N_{f}=1\right) $ whose graph is
given fig \ref{abc}-(b); and for a comment on the flavor charges of the
monopole and the dyon.

\subsection{Weak coupling chamber}

Extending the group theoretical approach described in previous section to
the present case,\ we can compute directly the BPS spectrum of this gauge
theory. This spectrum is given by the infinite set%
\begin{equation}
\begin{tabular}{lll}
$\gamma _{1}^{\left( n\right) },$ & $\gamma _{2}^{\left( n\right) },$ & $%
\gamma _{3}^{\left( n\right) }$ \\ 
&  &  \\ 
$\gamma _{1}^{\prime \left( n\right) },$ & $\gamma _{2}^{\prime \left(
n\right) },$ & $\gamma _{3}^{\prime \left( n\right) }$%
\end{tabular}
\label{spe}
\end{equation}%
with $\{ \gamma _{i}^{\left( n\right) }\},$ $\{ \gamma _{i}^{\prime \left(
n\right) }\}$, $n\in \mathbb{N}$; describing respectively the left and the
right sectors. These sectors are obtained separately by performing left
mutations and then right ones. As in eq(\ref{2K}); it turns out that the BPS
spectrum is given as well by several sequences. More precisely, we have 
\emph{6} families given by the classes of $\mathbb{Z}/6\mathbb{Z}$; and so
we have to distinguish different sub-chambers according to of the values of
the integer $n$. We have:

\  \  \  \ 

$\left( i\right) $ \emph{the class }$\emph{n=6k}$ \  \newline
This sequence describe a BPS sub-chamber of the weak coupling limit; the BPS
and anti-BPS states in this sub-chamber are obtained by performing left and
right quiver mutations. Under \emph{6k} successive elementary left mutations
of the primitive quiver $\mathfrak{Q}_{0}$, we obtain 
\begin{equation}
\begin{tabular}{lll}
$\gamma _{1}^{\left( 6k\right) }$ & $=\left( 1+3k\right) \gamma
_{1}+3k\left( \gamma _{2}+\gamma _{3}\right) $ & $=\left( 6k,1,\frac{1}{2}%
\right) $ \\ 
$\gamma _{2}^{\left( 6k\right) }$ & $=\left( 1-3k\right) \gamma
_{2}-3k\left( \gamma _{1}+\gamma _{3}\right) $ & $=\left( 1-6k,-1,\frac{1}{2}%
\right) $ \\ 
$\gamma _{3}^{\left( 6k\right) }$ & $=\gamma _{3}$ & $=\left( 1,0,-1\right) $%
\end{tabular}
\label{1}
\end{equation}%
with k a positive integer. Similarly under \emph{6k} successive elementary
right mutations; we get%
\begin{eqnarray}
&&%
\begin{tabular}{lll}
$\gamma _{1}^{\prime \left( 6k\right) }$ & $=\left( 1-3k\right) \gamma
_{1}-3k\left( \gamma _{2}+\gamma _{3}\right) $ & $=\left( -6k,1,\frac{1}{2}%
\right) $ \\ 
$\gamma _{2}^{\prime \left( 6k\right) }$ & $=\left( 1+3k\right) \gamma
_{2}+3k\left( \gamma _{1}+\gamma _{3}\right) $ & $=\left( 6k+1,-1,\frac{1}{2}%
\right) $ \\ 
$\gamma _{3}^{\prime \left( 6k\right) }$ & $=\gamma _{3}$ & $=\left(
1,0,-1\right) $%
\end{tabular}
\label{2} \\
&&  \notag
\end{eqnarray}%
Notice that this sub-chamber contains the elementary BPS states for $k=0$;
and the two sets (\ref{1}-\ref{2}) are related by the change $%
k\leftrightarrow -k$. Notice also that the sub-chamber is not CPT invariant.

\  \  \  \  \  \  \  \  \  \  \  \  \ 

$\left( ii\right) $ \emph{the class }$\emph{n=6k+1}$\newline
Under \emph{6k+1} successive elementary left mutations of $\mathfrak{Q}_{0}$%
, we have%
\begin{equation}
\begin{tabular}{lll}
$\gamma _{1}^{\left( 6k+1\right) }$ & $=-\left( 3k+1\right) \gamma
_{1}-3k\left( \gamma _{2}+\gamma _{3}\right) $ & $=\left( {\small -6k,-1,-}%
\frac{1}{2}\right) $ \\ 
$\gamma _{2}^{\left( 6k+1\right) }$ & $=\gamma _{1}+\gamma _{2}$ & $=\left( 
{\small 1,0,1}\right) $ \\ 
$\gamma _{3}^{\left( 6k+1\right) }$ & $=\left( 3k+1\right) \left( \gamma
_{1}+\gamma _{3}\right) +3k\gamma _{2}$ & $=\left( {\small 6k+1,1,-}\frac{1}{%
2}\right) $%
\end{tabular}%
\end{equation}%
and under \emph{6k+1} successive right ones, we get moreover%
\begin{equation}
\begin{tabular}{lll}
$\gamma _{1}^{\prime \left( 6k+1\right) }$ & $=-\left( 3k+1\right) \gamma
_{1}-3k\left( \gamma _{2}+\gamma _{3}\right) $ & $=\left( -6k,-1,-\frac{1}{2}%
\right) $ \\ 
$\gamma _{2}^{\prime \left( 6k+1\right) }$ & $=\left( 3k+1\right) \left(
\gamma _{1}+\gamma _{2}\right) +3k\gamma _{3}$ & $=\left( 6k+1,0,1\right) $
\\ 
$\gamma _{3}^{\prime \left( 6k+1\right) }$ & $=\gamma _{1}+\gamma _{3}$ & $%
=\left( 1,1,-\frac{1}{2}\right) $%
\end{tabular}%
\end{equation}

\  \  \  \  \  \  \  \  \  \  \  \  \  \  \  \  \  \  \  \ 

$\left( iii\right) $ \emph{the class }$\emph{n=6k+2}$\newline
The $\left( \emph{6k+2}\right) $ left mutations of $\mathfrak{Q}_{0}$ give%
\begin{equation}
\begin{tabular}{lll}
$\gamma _{1}^{\left( 6k+2\right) }$ & $=\gamma _{3}$ & $=\left(
1,0,-1\right) $ \\ 
$\gamma _{2}^{\left( 6k+2\right) }$ & $=\left( 3k+2\right) \gamma
_{1}+\left( 3k+1\right) \left( \gamma _{2}+\gamma _{3}\right) $ & $=\left(
6k+2,1,\frac{1}{2}\right) $ \\ 
$\gamma _{3}^{\left( 6k+2\right) }$ & $=-\left( 3k+1\right) \left( \gamma
_{1}+\gamma _{3}\right) -3k\gamma _{2}$ & $=\left( -6k-1,-1,\frac{1}{2}%
\right) $%
\end{tabular}%
\end{equation}%
and the $\left( \emph{6k+2}\right) $ right ones to%
\begin{equation}
\begin{tabular}{lll}
$\gamma _{1}^{\prime \left( 6k+2\right) }$ & $=-\left( 3k+1\right) \left(
\gamma _{1}+\gamma _{3}\right) -3k\gamma _{2}$ & $=\left( -6k-1,-1,\frac{1}{2%
}\right) $ \\ 
$\gamma _{2}^{\prime \left( 6k+2\right) }$ & $=\left( 3k+1\right) \left(
\gamma _{1}+\gamma _{2}\right) +\left( 3k+2\right) \gamma _{3}$ & $=\left(
6k+3,0,-1\right) $ \\ 
$\gamma _{3}^{\prime \left( 6k+2\right) }$ & $=\gamma _{1}$ & $=\left( 0,1,%
\frac{1}{2}\right) $%
\end{tabular}%
\end{equation}

\  \  \  \  \  \  \  \  \  \  \  \  \  \  \  \  \ 

$\left( iv\right) $ \emph{the class }$\emph{n=6k+3}$\newline
Left mutations of $\mathfrak{Q}_{0}$ give%
\begin{equation}
\begin{tabular}{lll}
$\gamma _{1}^{\left( 6k+3\right) }$ & $=\left( 3k+2\right) \left( \gamma
_{1}+\gamma _{3}\right) +\left( 3k+1\right) \gamma _{2}$ & $=\left( {\small %
6k+3,1,}\frac{-1}{2}\right) $ \\ 
$\gamma _{2}^{\left( 6k+3\right) }$ & $=-\left( 3k+2\right) \gamma
_{1}-\left( 3k+1\right) \left( \gamma _{2}+\gamma _{3}\right) $ & $=\left( 
{\small -6k-2,-1,}\frac{-1}{2}\right) $ \\ 
$\gamma _{3}^{\left( 6k+3\right) }$ & $=\gamma _{1}+\gamma _{2}$ & $=\left( 
{\small 1,0,1}\right) $%
\end{tabular}%
\end{equation}%
and the right ones lead to 
\begin{equation}
\begin{tabular}{lll}
$\gamma _{1}^{\prime \left( 6k+3\right) }$ & $=-\left( 3k+1\right) \left(
\gamma _{1}+\gamma _{3}\right) -\left( 3k+2\right) \gamma _{2}$ & $=\left(
-6k-3,1,-\frac{1}{2}\right) $ \\ 
$\gamma _{2}^{\prime \left( 6k+3\right) }$ & $=\left( 3k+1\right) \gamma
_{1}+\left( 3k+2\right) \left( \gamma _{2}+\gamma _{3}\right) $ & $=\left(
6k+4,-1,-\frac{1}{2}\right) $ \\ 
$\gamma _{3}^{\prime \left( 6k+3\right) }$ & $=\gamma _{1}+\gamma _{2}$ & $%
=\left( 1,0,1\right) $%
\end{tabular}%
\end{equation}

$\  \  \  \  \  \  \  \  \ $

$\left( v\right) $ \emph{the class }$\emph{n=6k+4}$\newline
In this sub-chamber, we have%
\begin{equation}
\begin{tabular}{lll}
$\gamma _{1}^{\left( 6k+4\right) }$ & $=-\left( 3k+2\right) \left( \gamma
_{1}+\gamma _{3}\right) -\left( 3k+1\right) \gamma _{2}$ & $=\left( -6k-3,-1,%
\frac{1}{2}\right) $ \\ 
$\gamma _{2}^{\left( 6k+4\right) }$ & $=\gamma _{3}$ & $=\left(
1,0,-1\right) $ \\ 
$\gamma _{3}^{\left( 6k+4\right) }$ & $=\left( 3k+3\right) \gamma
_{1}+\left( 3k+2\right) \left( \gamma _{2}+\gamma _{3}\right) $ & $=\left(
6k+4,1,\frac{1}{2}\right) $%
\end{tabular}%
\end{equation}%
and%
\begin{equation}
\begin{tabular}{lll}
$\gamma _{1}^{\prime \left( 6k+4\right) }$ & $=-\left( 3k+2\right) \left(
\gamma _{1}+\gamma _{2}\right) -\left( 3k+1\right) \gamma _{3}$ & $=\left( 
{\small -6k-3,0,-1}\right) $ \\ 
$\gamma _{2}^{\prime \left( 6k+4\right) }$ & $=\left( 3k+3\right) \gamma
_{1}+\left( 3k+2\right) \left( \gamma _{2}+\gamma _{3}\right) $ & $=\left( 
{\small 6k+4,1,}\frac{1}{2}\right) $ \\ 
$\gamma _{3}^{\prime \left( 6k+4\right) }$ & $=\gamma _{2}$ & $=\left( 
{\small 1,-1,}\frac{1}{2}\right) $%
\end{tabular}%
\end{equation}

\  \  \  \  \  \  \  \  \  \  \  \  \  \  \  \  \  \  \  \  \ 

$\left( vi\right) $ \emph{the class }$\emph{n=6k}+5$\newline
Under \emph{6k+5} left mutations of $\mathfrak{Q}_{0}$, we have%
\begin{equation}
\begin{tabular}{lll}
$\gamma _{1}^{\left( 6k+5\right) }$ & $=\gamma _{1}+\gamma _{2}$ & $=\left( 
{\small 1,0,1}\right) $ \\ 
$\gamma _{2}^{\left( 6k+5\right) }$ & $=\left( 3k+3\right) \left( \gamma
_{1}+\gamma _{3}\right) +\left( 3k+2\right) \gamma _{2}$ & $=\left( {\small %
6k+5,1,}\frac{-1}{2}\right) $ \\ 
$\gamma _{3}^{\left( 6k+5\right) }$ & $=-\left( 3k+3\right) \gamma
_{1}-\left( 3k+2\right) \left( \gamma _{2}+\gamma _{3}\right) $ & $=\left( 
{\small -6k-4,-1,}\frac{-1}{2}\right) $%
\end{tabular}%
\end{equation}%
and for the \emph{6k+5} right ones, we obtain%
\begin{equation}
\begin{tabular}{lll}
$\gamma _{1}^{\prime \left( 6k+5\right) }$ & $=-\left( 3k+2\right) \left(
\gamma _{1}+\gamma _{2}\right) -\left( 3k+3\right) \gamma _{3}$ & $=\left(
-6k-5,0,1\right) $ \\ 
$\gamma _{2}^{\prime \left( 6k+5\right) }$ & $=\left( 3k+3\right) \left(
\gamma _{1}+\gamma _{3}\right) +\left( 3k+2\right) \gamma _{2}$ & $=\left(
6k+5,1,-\frac{1}{2}\right) $ \\ 
$\gamma _{3}^{\prime \left( 6k+5\right) }$ & $=\gamma _{2}+\gamma _{3}$ & $%
=\left( 2,-1,-\frac{1}{2}\right) $%
\end{tabular}%
\end{equation}%
Below, we derive this BPS spectrum by using the quiver mutation group.

\subsection{Deriving the BPS spectrum}

Here we build the above BPS spectrum by using the group of quiver mutations.
The EM charges of the 3 elementary BPS particles $\gamma _{1},$ $\gamma
_{2}, $ $\gamma _{3}$ are as in (\ref{EM}); they generate the 3-dimensional
electric-magnetic lattice $\Gamma _{3}$.\ These elementary BPS states play a
crucial role in our analysis as they form the nodes of the primitive quiver $%
\mathfrak{Q}_{0}$ given by fig \ref{abc}-(b). Recall that this quiver is one
of the two basic object in the group theoretical approach; the other basic
object is given by the set of mutation matrices $\mathcal{M}_{m}$.\textrm{\ }%
To get the BPS states of the weak coupling chamber of this supersymmetric
gauge theory, we proceed in three steps as follows: (\textbf{1}) Introduce
the vector $\Upsilon ^{\left( 0\right) }$\textrm{\ }and the intersection
matrix $\mathcal{J}^{0}=\Upsilon ^{0}\circ \Upsilon ^{0}$ describing $%
\mathfrak{Q}_{0}$; these are given by%
\begin{equation}
\Upsilon ^{0}=\left( 
\begin{array}{c}
\gamma _{1} \\ 
\gamma _{2} \\ 
\gamma _{3}%
\end{array}%
\right) ,\qquad \mathcal{J}_{ij}^{0}=\left( 
\begin{array}{ccc}
0 & -1 & -1 \\ 
1 & 0 & 1 \\ 
1 & -1 & 0%
\end{array}%
\right)
\end{equation}%
(\textbf{2}) Perform successively the 3 following basic reflections 
\begin{eqnarray}
&&%
\begin{tabular}{lll}
$A=\left( 
\begin{array}{ccc}
-1 & 0 & 0 \\ 
1 & 1 & 0 \\ 
1 & 0 & 1%
\end{array}%
\right) ,$ & $B=\left( 
\begin{array}{ccc}
1 & 0 & 1 \\ 
0 & 1 & 1 \\ 
0 & 0 & -1%
\end{array}%
\right) ,$ & $C=\left( 
\begin{array}{ccc}
1 & 1 & 0 \\ 
0 & -1 & 0 \\ 
0 & 1 & 1%
\end{array}%
\right) $%
\end{tabular}
\label{A} \\
&&  \notag
\end{eqnarray}%
leading respectively to the 3 following mutated quivers $\mathfrak{Q}_{1}=A%
\mathfrak{Q}_{0},$ $\mathfrak{Q}_{2}=BA\mathfrak{Q}_{0}$, $\mathfrak{Q}%
_{3}=CBA\mathfrak{Q}_{0}$. Like for the primitive $\mathfrak{Q}_{0}$, these
mutated quivers are described by the vectors $\Upsilon ^{\left( n\right) }$
with $n=1,2,3;$ and the intersection matrices $\mathcal{J}_{ij}^{\left(
n\right) }$; we denote these objects as%
\begin{equation}
\Upsilon ^{\left( n\right) }=\left( 
\begin{array}{c}
\gamma _{1}^{\left( n\right) } \\ 
\gamma _{2}^{\left( n\right) } \\ 
\gamma _{3}^{\left( n\right) }%
\end{array}%
\right) ,\qquad \mathcal{J}_{ij}^{\left( n\right) }=\gamma _{i}^{\left(
n\right) }\circ \gamma _{j}^{\left( n\right) }  \label{qn}
\end{equation}%
Now, using the following convention notation that will be justified later on%
\begin{equation}
\begin{tabular}{lll}
$\mathcal{L}_{1}=A,$ & $\mathcal{L}_{2}=BA,$ & $\mathcal{L}_{3}=CBA,$ \\ 
&  &  \\ 
$\mathcal{R}_{1}=A,$ & $\mathcal{R}_{2}=AB,$ & $\mathcal{R}_{3}=ABC,$%
\end{tabular}%
\end{equation}%
one can express (\ref{qn}) in the following form 
\begin{equation}
\begin{tabular}{lll}
$\Upsilon ^{\left( n\right) }=\mathcal{L}_{n}\Upsilon ^{0}$ & , & $\mathcal{J%
}^{\left( n\right) }=\mathcal{L}_{n}\mathcal{J}_{ij}^{0}\mathcal{L}_{n}^{T}$
\\ 
&  &  \\ 
$\Upsilon ^{\prime \left( n\right) }=\mathcal{R}_{n}\Upsilon ^{0}$ & , & $%
\mathcal{J}^{\prime \left( n\right) }=\mathcal{R}_{n}\mathcal{J}_{ij}^{0}%
\mathcal{R}_{n}^{T}$%
\end{tabular}%
\end{equation}%
showing that all data about quiver mutations are captured by the $\mathcal{L}%
_{n}$'s and the $\mathcal{R}_{n}$'s. Notice that being reflections, we have%
\begin{equation}
\begin{tabular}{llll}
$A^{2}$ & $=B^{2}$ & $=C^{2}$ & $=I_{id}$ \\ 
$\det A$ & $=\det B$ & $=\det C$ & $=-1$%
\end{tabular}
\label{ref}
\end{equation}%
We also have 
\begin{eqnarray}
&&%
\begin{tabular}{lll}
$\mathcal{L}_{2}=\left( 
\begin{array}{ccc}
0 & 0 & 1 \\ 
2 & 1 & 1 \\ 
-1 & 0 & -1%
\end{array}%
\right) $ & , & $\mathcal{R}_{2}=\left( 
\begin{array}{ccc}
-1 & 0 & -1 \\ 
1 & 1 & 2 \\ 
1 & 0 & 0%
\end{array}%
\right) $ \\ 
&  &  \\ 
$\mathcal{L}_{3}=\left( 
\begin{array}{ccc}
2 & 1 & 2 \\ 
-2 & -1 & -1 \\ 
1 & 1 & 0%
\end{array}%
\right) $ & , & $\mathcal{R}_{3}=\left( 
\begin{array}{ccc}
-1 & -2 & -1 \\ 
1 & 2 & 2 \\ 
1 & 1 & 0%
\end{array}%
\right) $%
\end{tabular}
\\
&&  \notag
\end{eqnarray}%
satisfying some properties that follow from (\ref{ref}). In fact these
properties are particular relations encoded by the group structure of the
infinite set $\mathcal{G}_{weaak}=\left \{ \mathcal{L}_{m},\mathcal{R}_{m};%
\text{ }m\in \mathbb{N}\right \} $ whose matrix representation will be given
later. (\textbf{3}) Use the above particular quiver mutations to generate
all others by distinguishing the two kinds of mutations:

$\left( a\right) $ \emph{Left mutations} generated by the $\mathcal{L}_{n}$%
's and act as 
\begin{equation}
\Upsilon ^{\left( 0\right) }\rightarrow \Upsilon ^{\left( 1\right)
}\rightarrow \Upsilon ^{\left( 2\right) }\rightarrow \Upsilon ^{\left(
3\right) }\rightarrow \Upsilon ^{\left( 4\right) }\rightarrow ...
\end{equation}%
with $\Upsilon ^{\left( n\right) }=\mathcal{L}_{n}\Upsilon ^{\left( 0\right)
}$ and $\mathcal{J}^{\left( n\right) }=\mathcal{L}_{n}\mathcal{J}%
_{ij}^{\left( 0\right) }\mathcal{L}_{n}^{T}$ for any positive integer.

$\left( b\right) $ \emph{Right mutations} generated by the $\mathcal{R}_{n}$%
's and operate like 
\begin{equation}
\Upsilon ^{\left( 0\right) }\rightarrow \Upsilon ^{\prime \left( 1\right)
}\rightarrow \Upsilon ^{\prime \left( 2\right) }\rightarrow \Upsilon
^{\prime \left( 3\right) }\rightarrow \Upsilon ^{\prime \left( 4\right)
}\rightarrow ...
\end{equation}%
with $\Upsilon ^{\prime \left( n\right) }=\mathcal{R}_{n}\Upsilon ^{\left(
0\right) }$ and $\mathcal{J}^{\prime \left( n\right) }=\mathcal{L}%
_{n}^{\prime }\mathcal{J}_{ij}^{\left( 0\right) }\mathcal{L}_{n}^{\prime T}$
for any positive integer.\  \  \newline
These sequences of quiver mutations combine together to form the following
infinite set%
\begin{eqnarray}
&&%
\begin{tabular}{llllllll}
\multicolumn{2}{l}{left sector} &  &  &  &  & \multicolumn{2}{l}{right sector
} \\ 
&  &  &  &  &  &  &  \\ 
$\mathcal{L}_{3n}$ & $=\left( CBA\right) ^{n}$ &  &  &  &  & $\mathcal{R}%
_{3n}$ & $=\left( ABC\right) ^{n}$ \\ 
&  &  &  &  &  &  &  \\ 
$\mathcal{L}_{3n+1}$ & $=A\mathcal{L}_{3n}$ &  &  &  &  & $\mathcal{R}%
_{3n+1} $ & $=\mathcal{R}_{3n}A$ \\ 
&  &  &  &  &  &  &  \\ 
$\mathcal{L}_{3n+2}$ & $=BA\mathcal{L}_{3n}$ &  &  &  &  & $\mathcal{R}%
_{3n+2}$ & $=\mathcal{R}_{3n}AB$%
\end{tabular}
\label{B} \\
&&  \notag
\end{eqnarray}%
\ that turns out to be an infinite discrete subgroup $\mathcal{G}_{weak}$ of 
$GL\left( 3,\mathbb{Z}\right) $. The matrix realization of the group $%
\mathcal{G}_{weak}$ gives exactly the electric-magnetic charges of the BPS
states of the weak coupling chamber of this supersymmetric gauge theory. Let
us compute explicitly this spectrum.

\subsection{Explicit computation of the BPS spectrum}

The study of the matrix representation of the quiver mutation group $%
\mathcal{G}_{weak}$ reveals that the explicit realization of the $\mathcal{L}%
_{m}$ and $\mathcal{R}_{m}$ matrices can split into $\mathbb{Z}/6\mathbb{Z}$
classes as follows:%
\begin{eqnarray}
&&%
\begin{tabular}{llllllll}
\multicolumn{2}{l}{left sector} &  &  &  &  & \multicolumn{2}{l}{right sector
} \\ 
$\mathcal{L}_{6k}$ & $=\left( \mathcal{L}_{3}\right) ^{2k}$ &  &  &  &  & $%
\mathcal{R}_{6k}$ & $=\left( \mathcal{R}_{3}\right) ^{2k}$ \\ 
$\mathcal{L}_{6k+1}$ & $=A\mathcal{L}_{6k}$ &  &  &  &  & $\mathcal{R}%
_{6k+1} $ & $=\mathcal{R}_{6k}A$ \\ 
$\mathcal{L}_{6k+2}$ & $=B\mathcal{L}_{6k+1}$ &  &  &  &  & $\mathcal{R}%
_{6k+2}$ & $=\mathcal{R}_{6k+1}B$ \\ 
$\mathcal{L}_{6k+3}$ & $=\left( \mathcal{L}_{3}\right) ^{2k+1}$ &  &  &  & 
& $\mathcal{R}_{6k+3}$ & $=\left( \mathcal{R}_{3}\right) ^{2k+1}$ \\ 
$\mathcal{L}_{6k+4}$ & $=A\mathcal{L}_{6k+3}$ &  &  &  &  & $\mathcal{R}%
_{6k+4}$ & $=\mathcal{R}_{6k+3}A$ \\ 
$\mathcal{L}_{6k+5}$ & $=B\mathcal{L}_{6k+4}$ &  &  &  &  & $\mathcal{R}%
_{6k+5}$ & $=\mathcal{R}_{6k+4}B$%
\end{tabular}
\label{C} \\
&&  \notag
\end{eqnarray}%
with $k$ a positive integer. Using the properties on the generators, one can
show that the set $\left \{ \mathcal{L}_{n},\mathcal{R}_{m};\text{ }n,m\in 
\mathbb{N}\right \} $ has a discrete group structure. Moreover, using (\ref%
{A},\ref{C}), we obtain:

\  \  \  \ 

$(i)$ \emph{Left mutations}\newline
These mutations involves 6 infinite discrete series related to each other by
the basic reflections. The first series is given by the following set of $%
3\times 3$ matrices,%
\begin{equation}
\mathcal{L}_{6k}=\left( 
\begin{array}{ccc}
1+3k & 3k & 3k \\ 
-3k & 1-3k & -3k \\ 
0 & 0 & 1%
\end{array}%
\right) ,\qquad k\in \mathbb{N}  \label{L2K}
\end{equation}%
containing the identity $\mathcal{L}_{0}=I_{id}$. Notice that the matrix
elements of these series have the property $\det \mathcal{L}_{6k}=1$ $%
\forall k$; but do not form a subgroup since the inverse of these matrices $%
\left( \mathcal{L}_{6k}\right) ^{-1}$ do not belong to the set $\left \{ 
\mathcal{L}_{6k},\text{ }k\in \mathbb{N}\right \} $. As we will show later
on, the inverse belong to the right mutations set; a property that explains
the need of both left and right mutations noticed in \textrm{\cite{1A,1B}}
to compute the full BPS spectrum. The other elements of the set of left
mutations are given by eqs(\ref{C}). Setting $\mathcal{B}=\mathcal{L}_{%
{\small 6k+1}}$, $\mathcal{C}=\mathcal{L}_{{\small 6k+2}}$, $\mathcal{D}=%
\mathcal{L}_{{\small 6k+3}}$, $\mathcal{E}=\mathcal{L}_{{\small 6k+4}}$, $%
\mathcal{F}=\mathcal{L}_{{\small 6k+5}}$, we have:%
\begin{equation}
\mathcal{B}=\left( 
\begin{array}{ccc}
-3k-1 & -3k & -3k \\ 
1 & 1 & 0 \\ 
3k+1 & 3k & 3k+1%
\end{array}%
\right)
\end{equation}%
and%
\begin{eqnarray*}
&&%
\begin{tabular}{ll}
$\mathcal{C}=\left( 
\begin{array}{ccc}
0 & 0 & 1 \\ 
3k+2 & 3k+1 & 3k+1 \\ 
-3k-1 & -3k & -3k-1%
\end{array}%
\right) $ &  \\ 
&  \\ 
$\mathcal{D}=\left( 
\begin{array}{ccc}
3k+2 & 3k+1 & 3k+2 \\ 
-3k-2 & -3k-1 & -3k-1 \\ 
1 & 1 & 0%
\end{array}%
\right) $ & 
\end{tabular}
\\
&&
\end{eqnarray*}%
as well as%
\begin{eqnarray*}
&&%
\begin{tabular}{l}
$\mathcal{E}=\left( 
\begin{array}{ccc}
-3k-2 & -3k-1 & -3k-2 \\ 
0 & 0 & 1 \\ 
3k+3 & 3k+2 & 3k+2%
\end{array}%
\right) $ \\ 
\\ 
$\mathcal{F}=\left( 
\begin{array}{ccc}
1 & 1 & 0 \\ 
3k+3 & 3k+2 & 3k+3 \\ 
-3k-3 & -3k-2 & -3k-2%
\end{array}%
\right) $%
\end{tabular}
\\
&&
\end{eqnarray*}%
with the properties $\det \mathcal{L}_{{\small 6k+1}}=$ $\det \mathcal{L}_{%
{\small 6k+3}}=$ $\mathcal{L}_{{\small 6k+5}}=-1$; and $\det \mathcal{L}_{%
{\small 6k+2}}=\mathcal{L}_{{\small 6k+4}}=1$.

\  \  \  \  \ 

$(ii)$ \emph{Right mutations}\newline
Right mutations, contributing to the building of the weak chamber of BPS
states of the supersymmetric $SU\left( 2\right) $ gauge theory with a quark
hypermultiplet, involve 6 infinite series related to each other by the basic
A, B, C reflections. The first series is given by the set $\left \{ \mathcal{%
R}_{6k},\ k\in \mathbb{N}\right \} $ with $\det \mathcal{R}_{6k}=1$; and $%
\mathcal{R}_{6k}$ exactly the inverse of $\mathcal{L}_{6k}$ for any positive
integer k. This property shows that $\left \{ \mathcal{L}_{6k},\mathcal{R}%
_{6k}\text{, }k\in \mathbb{N}\right \} $ form a subgroup of the full set of
quiver mutations $\mathcal{G}_{weak}$. The full set of right mutations is
given by%
\begin{equation}
\begin{tabular}{lll}
$\mathcal{A}^{\prime }=\mathcal{R}_{6k},$ & $\mathcal{C}^{\prime }=\mathcal{R%
}_{6k+2},$ & $\mathcal{E}^{\prime }=\mathcal{R}_{6k+4}$ \\ 
$\mathcal{B}^{\prime }=\mathcal{R}_{6k+1},$ & $\mathcal{D}^{\prime }=%
\mathcal{R}_{6k+3},$ & $\mathcal{F}^{\prime }=\mathcal{R}_{6k+5}$%
\end{tabular}%
\end{equation}%
The explicit expression of these matrices are as follows 
\begin{eqnarray*}
&&%
\begin{tabular}{ll}
$\mathcal{A}^{\prime }=\left( 
\begin{array}{ccc}
1-3k & -3k & -3k \\ 
3k & 1+3k & 3k \\ 
0 & 0 & 1%
\end{array}%
\right) $ &  \\ 
&  \\ 
$\mathcal{B}^{\prime }=\left( 
\begin{array}{ccc}
-3k-1 & -3k & -3k \\ 
3k+1 & 3k+1 & 3k \\ 
1 & 0 & 1%
\end{array}%
\right) $ &  \\ 
&  \\ 
$\mathcal{C}^{\prime }=\left( 
\begin{array}{ccc}
-3k-1 & -3k & -3k-1 \\ 
3k+1 & 3k+1 & 3k+2 \\ 
1 & 0 & 0%
\end{array}%
\right) $ & 
\end{tabular}
\\
&&
\end{eqnarray*}%
and%
\begin{eqnarray*}
&&%
\begin{tabular}{l}
$\mathcal{F}^{\prime }=\left( 
\begin{array}{ccc}
-3k-2 & -3k-2 & -3k-3 \\ 
3k+3 & 3k+2 & 3k+3 \\ 
0 & 1 & 1%
\end{array}%
\right) $ \\ 
\\ 
$\mathcal{E}^{\prime }=\left( 
\begin{array}{ccc}
-3k-2 & -3k-2 & -3k-1 \\ 
3k+3 & 3k+2 & 3k+2 \\ 
0 & 1 & 0%
\end{array}%
\right) $ \\ 
\\ 
$\mathcal{D}^{\prime }=\left( 
\begin{array}{ccc}
-3k-1 & -3k-2 & -3k-1 \\ 
3k+1 & 3k+2 & 3k+2 \\ 
1 & 1 & 0%
\end{array}%
\right) $%
\end{tabular}
\\
&&
\end{eqnarray*}%
The rows of the matrices $\mathcal{L}_{6m}$ and $\mathcal{R}_{m}$ give
exactly the electric-magnetic charges $\gamma _{1}^{\left( m\right) },$ $%
\gamma _{2}^{\left( m\right) },$ $\gamma _{3}^{\left( m\right) }$ of the BPS
states of the weak coupling chamber. In the end of this section, notice that
like in the pure SU$\left( 2\right) $ gauge theory, here also the number of
the BPS states of the weak coupling chamber grows linearly with respect to
the integer k; so by defining the infinite the limit of the mutation
matrices like 
\begin{equation}
\mathcal{M}_{\infty }=\lim_{m\rightarrow \infty }\frac{1}{m}\mathcal{M}_{m}
\end{equation}%
we end with singular matrices that are associated with the BPS gauge
particles. For instance, we have for the mutation matrix (\ref{L2K}) the
following singular limit: 
\begin{equation}
\begin{tabular}{ll}
$\lim_{k\rightarrow \infty }\frac{1}{3k}\mathcal{L}_{6k}=\left( 
\begin{array}{ccc}
1 & 1 & 1 \\ 
-1 & -1 & -1 \\ 
0 & 0 & 1%
\end{array}%
\right) ,$ & $\det \mathcal{L}_{\infty }=0$%
\end{tabular}%
\end{equation}%
describing, in addition to the hypermultiplet with charge $\gamma _{3}$, two
BPS vector particles with electric-magnetic charges $\pm w$ with $w=\gamma
_{1}+\gamma _{2}+\gamma _{3}$ which, by help of (\ref{EM}), is equal to $%
\left( 2,0,0\right) $.

\section{$SU\left( 2\right) $ model with two hypermultiplets}

\begin{figure}[tbph]
\begin{center}
\hspace{0cm} \includegraphics[width=15cm]{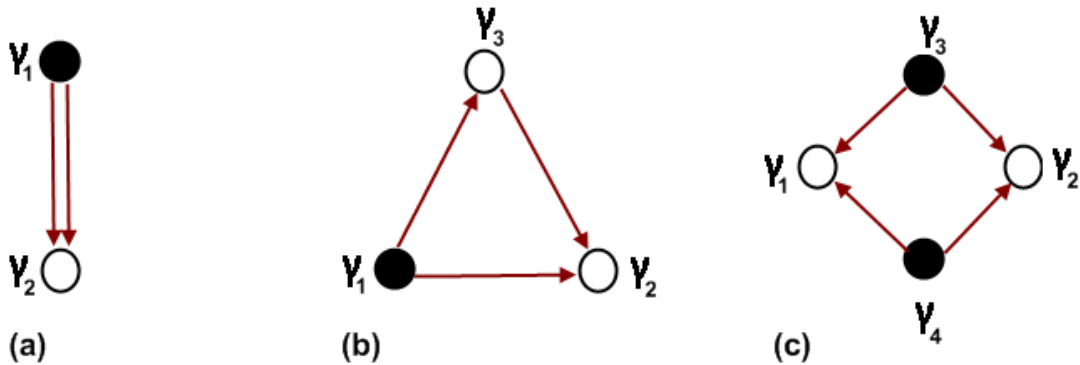}
\end{center}
\par
\vspace{-1 cm}
\caption{{\protect \small Primitive quivers: }(a) $\mathcal{N}=2$ 
{\protect \small pure SU}$\left( {\protect \small 2}\right) $ {\protect \small %
gauge model, (b) adding one quark hypermultiplet, (c) adding\ two
hypermultiplets}.}
\label{abc}
\end{figure}

The primitive quiver $\mathfrak{Q}_{0}$ of the $\mathcal{N}=2$
supersymmetric SU$\left( 2\right) $ gauge model with two quark
hypermultiplets involves 4 particles with electric-magnetic $\gamma _{1},$ $%
\gamma _{2},$ $\gamma _{3},$ $\gamma _{4}$ as depicted on fig \ref{abc}-(c).
The electric-magnetic charges are as follows%
\begin{equation}
\begin{tabular}{llll}
$\gamma _{1}=\left( 
\begin{array}{c}
1 \\ 
-1 \\ 
0 \\ 
-\frac{1}{2}%
\end{array}%
\right) ,$ & $\gamma _{2}=\left( 
\begin{array}{c}
1 \\ 
-1 \\ 
0 \\ 
\frac{1}{2}%
\end{array}%
\right) ,$ & $\gamma _{3}=\left( 
\begin{array}{c}
0 \\ 
1 \\ 
\frac{1}{2} \\ 
0%
\end{array}%
\right) ,$ & $\gamma _{4}=\left( 
\begin{array}{c}
0 \\ 
1 \\ 
-\frac{1}{2} \\ 
0%
\end{array}%
\right) $%
\end{tabular}
\label{me}
\end{equation}%
and the intersection matrix $\mathcal{J}_{ij}^{0}$, computed by help of the
electric-magnetic products of the $\gamma _{i}$'s, is as given below. The
extra third and fourth entries of these EM charge vectors refer to the $%
U_{f}\left( 1\right) \times U_{f}^{\prime }\left( 1\right) $ charges of the
flavor symmetry of the gauge model namely $SO_{f}\left( 4\right) \sim
SU_{f}\left( 2\right) \times SU_{f}^{\prime }\left( 2\right) $. In our group
theoretical set up, the primitive quiver is described by the pair $\left(
\Upsilon ^{0},\mathcal{J}_{ij}^{0}\right) $ with 
\begin{equation}
\Upsilon ^{0}=\left( 
\begin{array}{c}
\gamma _{1} \\ 
\gamma _{2} \\ 
\gamma _{3} \\ 
\gamma _{4}%
\end{array}%
\right) ,\qquad \mathcal{J}_{ij}^{0}=\left( 
\begin{array}{cccc}
0 & 0 & 1 & 1 \\ 
0 & 0 & 1 & 1 \\ 
-1 & -1 & 0 & 0 \\ 
-1 & -1 & 0 & 0%
\end{array}%
\right)
\end{equation}%
The BPS spectrum of the weak coupling chamber of this supersymmetric gauge
theory with two quark hypermultiplets can be obtained by extending the group
theoretical method used in the previous sections. In this case also, the
quiver mutation group of the weak coupling chamber is an infinite discrete
set $\mathcal{G}_{weak}=\left \{ \mathcal{L}_{m},\mathcal{R}_{m}\text{; }%
m\in \mathbb{N}\right \} $ generated by the two following reflections 
\begin{equation}
\begin{tabular}{ll}
$A=\left( 
\begin{array}{cccc}
1 & 0 & 1 & 1 \\ 
0 & 1 & 1 & 1 \\ 
0 & 0 & -1 & 0 \\ 
0 & 0 & 0 & -1%
\end{array}%
\right) ,$ & $B=\left( 
\begin{array}{cccc}
-1 & 0 & 0 & 0 \\ 
0 & -1 & 0 & 0 \\ 
1 & 1 & 1 & 0 \\ 
1 & 1 & 0 & 1%
\end{array}%
\right) $%
\end{tabular}
\label{ld}
\end{equation}%
satisfying $A^{2}=B^{2}=I_{id}$. As in previous examples, here also the weak
coupling chamber has several sub-chambers given by the four sequences $%
\left
\{ \mathcal{L}_{4k+q};\mathcal{R}_{4k+q}\right \} $ with $q=0,1,2,3$.
The general terms of these sequences are as follows,%
\begin{eqnarray}
&&%
\begin{tabular}{llll}
$\mathcal{L}_{4k}=\left( BA\right) ^{2k},$ & $\mathcal{L}_{4k+1}=A\mathcal{L}%
_{4k},$ & $\mathcal{L}_{4k+2}=B\mathcal{L}_{4k+1}$ & $\mathcal{L}_{4k+3}=A%
\mathcal{L}_{4k+2}$ \\ 
&  &  &  \\ 
$\mathcal{R}_{4k}=\left( AB\right) ^{2k},$ & $\mathcal{R}_{4k+1}=B\mathcal{R}%
_{4k},$ & $\mathcal{R}_{4k+2}=A\mathcal{R}_{4k+1}$ & $\mathcal{R}_{4k+3}=B%
\mathcal{R}_{4k+2}$%
\end{tabular}
\label{dl} \\
&&  \notag
\end{eqnarray}%
Clearly this infinite set $\mathcal{G}_{weak}$ has a discrete group
structure with identity element given by the 4$\times $4 identity matrix; in
fact $\mathcal{G}_{weak}$ is a particular subgroup of $GL\left( 4,\mathbb{Z}%
\right) $. Using the matrix expressions of the reflections A and B, we
obtain for the $\mathcal{L}_{4k}$ left mutations%
\begin{equation}
\mathcal{L}_{4k}=\left( 
\begin{array}{cccc}
1-2k & -2k & -2k & -2k \\ 
-2k & 1-2k & -2k & -2k \\ 
2k & 2k & 2k+1 & 2k \\ 
2k & 2k & 2k & 2k+1%
\end{array}%
\right)
\end{equation}%
and for the corresponding right mutations 
\begin{equation}
\mathcal{R}_{4k}=\left( 
\begin{array}{cccc}
2k+1 & 2k & 2k & 2k \\ 
2k & 2k+1 & 2k & 2k \\ 
-2k & -2k & 1-2k & -2k \\ 
-2k & -2k & -2k & 1-2k%
\end{array}%
\right)
\end{equation}%
which is exactly $\left( \mathcal{L}_{4k}\right) ^{-1}$. The expressions of
the other mutations matrices are directly learnt from (\ref{dl}-\ref{ld}).
After doing these computations, we can determine the electric-magnetic
charges of the BPS and anti-BPS states from the rows of the various mutation
matrices. For BPS states, we find the following spectrum%
\begin{equation}
\begin{tabular}{lll}
$2k\gamma _{1}+\left( 2k+1\right) \left( \gamma _{2}+\gamma _{3}+\gamma
_{4}\right) $ & , & $\left( 2k+1\right) \gamma _{1}+2k\left( \gamma
_{2}+\gamma _{3}+\gamma _{4}\right) $ \\ 
$2k\gamma _{2}+\left( 2k+1\right) \left( \gamma _{1}+\gamma _{3}+\gamma
_{4}\right) $ & , & $\left( 2k+1\right) \gamma _{2}+2k\left( \gamma
_{1}+\gamma _{3}+\gamma _{4}\right) $ \\ 
$\left( 2k+1\right) \gamma _{1}+\left( 2k+2\right) \left( \gamma _{2}+\gamma
_{3}+\gamma _{4}\right) $ & , & $\left( 2k+1\right) \gamma _{3}+2k\left(
\gamma _{1}+\gamma _{2}+\gamma _{4}\right) $ \\ 
$\left( 2k+1\right) \gamma _{2}+\left( 2k+2\right) \left( \gamma _{1}+\gamma
_{3}+\gamma _{4}\right) $ & , & $\left( 2k+1\right) \gamma _{4}+2k\left(
\gamma _{1}+\gamma _{2}+\gamma _{3}\right) $%
\end{tabular}%
\end{equation}%
and%
\begin{eqnarray}
&&%
\begin{tabular}{l}
$\left( 2k+2\right) \gamma _{1}+\left( 2k+1\right) \left( \gamma _{2}+\gamma
_{3}+\gamma _{4}\right) $ \\ 
$\left( 2k+2\right) \gamma _{2}+\left( 2k+1\right) \left( \gamma _{1}+\gamma
_{3}+\gamma _{4}\right) $ \\ 
$\left( 2k+2\right) \gamma _{3}+\left( 2k+1\right) \left( \gamma _{1}+\gamma
_{2}+\gamma _{4}\right) $ \\ 
$\left( 2k+2\right) \gamma _{4}+\left( 2k+1\right) \left( \gamma _{1}+\gamma
_{2}+\gamma _{3}\right) $%
\end{tabular}
\\
&&
\end{eqnarray}%
The corresponding anti-BPS states have opposite electric-magnetic charges.
In the end of this study, notice that the BPS spectrum grows linearly with
integer k. For the limit $k\rightarrow \infty $, one ends with singular
matrices as shown on the following example%
\begin{equation}
\begin{tabular}{ll}
$\lim_{k\rightarrow \infty }\frac{1}{2k}\mathcal{L}_{4k}=\left( 
\begin{array}{cccc}
-1 & -1 & -1 & -1 \\ 
-1 & -1 & -1 & -1 \\ 
1 & 1 & 1 & 1 \\ 
1 & 1 & 1 & 1%
\end{array}%
\right) ,$ & $\det \mathcal{L}_{\infty }=0$%
\end{tabular}%
\end{equation}%
This singular limit describes BPS vector states with electric magnetic
charge vectors as $\pm w$ with $w=\gamma _{1}+$ $\gamma _{2}+\gamma
_{3}+\gamma _{4}$; which by help of (\ref{me}) is noting but the charge $%
\left( 2,0,0,0\right) $ of the massive vector bosons of the spontaneously
broken SU$\left( 2\right) $ gauge symmetry.

\section{Conclusion and comment}

In this paper, we have studied the weak coupling chamber of some models of  $%
\mathcal{N}=2$\ supersymmetric SU$\left( 2\right) $ gauge theory by using
the symmetry structure of the set of quiver mutations of the BPS quiver
approach of \textrm{\cite{1A,1B}}. Generally, the quiver mutations\ of $%
\mathcal{N}=2$\ supersymmetric theories are discrete symmetries of BPS
chambers; and may form either a finite set or an infinite one. Finite quiver
mutations concern the strong coupling chambers of supersymmetric QFT's as
shown in \textrm{\cite{3B}; }these discrete mutations turn out to be
isomorphic to a class of (dihedral) subgroups of the well known finite
Coxeter groups. The latters are used in building of the Dynkin diagrams of
finite dimensional Lie algebra of the gauge symmetries. Infinite mutations
are symmetries of weak coupling chambers of supersymmetric QFTs. The present
study has been done for the case of SU$\left( 2\right) $ gauge theory
without and with fundamental matter; but this construction can be also
applied for $\mathcal{N}=2$ supersymmetric QFTs with higher dimensional
gauge symmetries.

\  \  \  \  \  \newline
\textbf{Acknowledgements}: {\small This research work is supported by
URAC09/CNRS.}


\begin{thebibliography}{99}
\bibitem{1A} Murad Alim, Sergio Cecotti, Clay Cordova, Sam Espahbodi, Ashwin
Rastogi, Cumrun Vafa, \emph{BPS Quivers and Spectra of Complete }$\mathcal{N}%
=2$\emph{\ Quantum Field Theories}, arXiv:1109.4941,

\bibitem{1B} Murad Alim, Sergio Cecotti, Clay Cordova, Sam Espahbodi, Ashwin
Rastogi, Cumrun Vafa, $\mathcal{N}$\emph{=2 Quantum Field Theories and Their
BPS Quivers}, arXiv:1112.3984,

\bibitem{1C} S. Cecotti and C. Vafa, Classification of complete $\mathcal{N}%
=2$ supersymmetric theories in 4dimensions, arXiv:1103.5832,

\bibitem{2A} N. Seiberg and E. Witten, \emph{Electric - magnetic duality,
monopole condensation, and confinement in N=2 supersymmetric Yang-Mills
theory}, Nucl.Phys. B426 (1994) 1952, arXiv:hep-th/9407087 [hep-th].

\bibitem{2B} N. Seiberg and E. Witten, \emph{Monopoles, duality and chiral
symmetry breaking in N=2 supersymmetric QCD}, Nucl.Phys. B431 (1994) 484550,
arXiv:hep-th/9408099,

\bibitem{2C} A. Klemm, W. Lerche, S. Yankielowicz, and S. Theisen. \emph{%
Simple singularities and} $\mathcal{N}=2$ \emph{supersymmetric Yang-Mills
theory}. Phys. Lett., B344:169--175, 1995, hep-th/9411048,

\bibitem{2D} Philip C. Argyres and Alon E. Faraggi. \emph{The vacuum
structure and spectrum of} $\mathcal{N}=2$ \emph{supersymmetric su(n) gauge
theory}. Phys. Rev. Lett., 74:3931--3934, 1995, hep-th/9411057,

\bibitem{3A} A. Bilal, \emph{Duality in} $\mathcal{N}=2$ \emph{SUSY Gauge
Theories: low-energy effective action and BPS spectra}, arXiv:hep-th/0106246,

\bibitem{A3} Michael Yu. Kuchiev, \emph{Charges of dyons in} $\mathcal{N}=2$ 
\emph{supersymmetric gauge theory}, Nucl.Phys.B803:113-134,2008,
arXiv:0805.1461,

\bibitem{3} Michael Yu. Kuchiev. \emph{Supersymmetric} $\mathcal{N}=2$ \emph{%
gauge theory with arbitrary gauge group}. Nucl.Phys., B838:331--357, 2010,
0907.2010.

\bibitem{3B} E.H Saidi, \emph{Mutations Symmetries in BPS quiver theory:
Building the BPS Spectra}, arXiv:1204.0395, Journal of High Energy Physics,
2012, Volume 2012, Number 8, 18.
\end{thebibliography}
\end{document}